# Dynamical properties of three terminal magnetic tunnel junctions: spintronics meets spin-orbitronics

R. Tomasello,[1] M. Carpentieri,[2] G. Finocchio[3]

[1]Department of Computer Science, Modeling, Electronics and System Science, University of Calabria, Rende (CS), Italy.

[2]Department of Electrical and Information Engineering, Politecnico of Bari, via E. Orabona 4, I-70125 Bari, Italy.

[3]Department of Electronic Engineering, Industrial Chemistry and Engineering, University of Messina, C.da di Dio, I-98166, Messina, Italy.

**Abstract:** This letter introduces a micromagnetic model able to characterize the magnetization dynamics in three terminal magnetic tunnel junctions, where the effects of spin-transfer torque and spin-orbit torque are taken into account. Our results predict that the possibility to separate electrically those two torque sources is very promising from a technological point of view for both next generation of nanoscale spintronic oscillators and microwave detectors. A scalable synchronization scheme based on the parallel connection of those three terminal devices is also proposed.

Corresponding author email: m.carpentieri@poliba.it
Telephone: +39.080.5963254

Experimental demonstrations of magnetization switching,[1, 2] domain wall motion[3] and persistent magnetization precession,[4] induced by an in-plane current injection in heavy metal/ferromagnetic/oxide trilayer, have drawn increasing interest to spin torques based on orbital-to-spin momentum transfer (SOT) from Rashba effect (RE) and spin-Hall effect (SHE).[5, 6, 7] Despite a huge number of experiments, a detailed theoretical analysis based on complete micromagnetic simulations to deeply understand the inhomogeneous magnetization processes is missing. This understanding is crucial from both fundamental and technological point of view. Particularly, in the design of the next generation of spintronic devices, together to the advantage to use SOT and especially the SHE (obtaining, in this way, spin-injection without the presence of a ferromagnetic polarizer layer), it will be essential to include the spin-transfer torque (STT) from polarized currents, in order to improve the efficiency and the dynamical properties of those devices. In this letter, we study a three terminal device, which efficiently couples spintronics with spin-orbitronics. The pioneering idea of that system has been introduced by *Liu et al*[8,9]. However, in that experimental work, the current injected via the third terminal was only used to control the interfacial perpendicular anisotropy maintaining the STT negligible. Differently, here we predict the behavior of that system when also the STT contribution is significant. The three terminal device is composed by a magnetic tunnel junction (MTJ) CoFeB(2)/MgO(1.2)/CoFeB(4)/Ta(5)/Ru(5) (thicknesses in nm) built over a Tantalum (Ta) strip (6000x1200x6 nm$^3$).[9] The CoFeB(2) and the CoFeB(4) act as free and pinned layer of the MTJ respectively. Fig. 1 shows a detailed sketch of the system. We introduce a Cartesian coordinate system where the *x*-axis is oriented along the larger dimension of the Tantalum strip, the *y*- and *z*-axis are related to its width and thickness respectively. There are many advantages to study this system. First of all, the magnetization precession is read via the tunneling magneto-resistive effect instead of the anisotropic magneto resistance (AMR).[10] In fact, the oscillator output power can reach the same order of magnitude of the "*state-of-the-art*" MTJs-based oscillators.[11, 12, 13] Secondly, it is possible to control the injection of two current densities: the in-plane $J_{Ta}$ in the Tantalum strip and the perpendicular $J_{MTJ}$ flowing into the MTJ-stack, achieving an additional degree of freedom in the control of the magnetization dynamics. In addition, the use of Ta/CoFe/MgO gives rise to a spin-Hall angle two times larger than the one in Pt/Co/AlO.[8]

Our findings are important both in the understanding the types of spatially-inhomogeneous dynamics, that can be excited in presence of SOTs, and for an optimized design of devices which couple spintronics and spin-orbitronics. The two main results of this letter are: (i) micromagnetic understanding of the dynamical properties of those oscillators in terms of oscillation frequency and spatial distribution of the excited modes, and (ii) a spintronic/spin-orbitronic synchronization scheme which can be used either to improve the properties of the oscillators (linewidth, output power) or to enhance the sensitivity of resonant microwave signal detectors.

A self-implemented *"state of the art"* micromagnetic solver has been used to numerically solve the Landau-Lifshitz-Gilbert equation,[14] which includes the STT from a spin-polarized current[15] and the SOT driven by SHE:[2, 16, 17]

$$\frac{d\mathbf{m}}{\gamma_0 M_S dt} = -\frac{1}{(1+\alpha^2)}\mathbf{m}\times\mathbf{h}_{EFF} - \frac{\alpha}{(1+\alpha^2)}\mathbf{m}\times\mathbf{m}\times\mathbf{h}_{EFF}$$
$$-\frac{d_J}{(1+\alpha^2)\gamma_0 M_S}\mathbf{m}\times\mathbf{m}\times\boldsymbol{\sigma} + \frac{\alpha d_J}{(1+\alpha^2)\gamma_0 M_S}\mathbf{m}\times\boldsymbol{\sigma} \quad (1)$$
$$+\frac{g}{|e|\gamma_0}\frac{|\mu_B|J_{MTJ}}{M_s^2 t}g_T(\mathbf{m},\mathbf{m_p})\left[\mathbf{m}\times(\mathbf{m}\times\mathbf{m_p})-q(V)(\mathbf{m}\times\mathbf{m_p})\right]$$

being $\mathbf{m}$, $\mathbf{h}_{EFF}$ and $\mathbf{m_p}$, the magnetization and the effective field of the CoFeB free layer and the magnetization of the polarizer (fixed along the –y direction). $g$ is the Landè factor, $\mu_B$ the Bohr magneton, $e$ the electron charge, $\gamma_0$ the gyromagnetic ratio, $\alpha$ the Gilbert damping, $M_s$ the saturation magnetization, and $t$ the thickness of the free layer. $J_{MTJ}$ is the current density flowing through the MTJ stack, $g_T(\mathbf{m},\mathbf{m_p}) = \frac{2\eta_T}{1+\eta_T^2 \mathbf{m}\cdot\mathbf{m_p}}$ characterizes the angular dependence of the spin-polarization function as computed by Slonczewski,[18, 19] $\eta_T$ is the polarization efficiency. $q(V)$ is a function which takes into account the voltage dependence of the field-like torque term in the

MTJ.[20, 21] The coefficient $d_J = \frac{\mu_B \alpha_H}{e M_S t} J_{Ta}$, being $\alpha_H$ the spin Hall angle (ratio between spin-current $J_{SHE}$ and $J_{Ta}$). **σ** is the direction of the $J_{SHE}$ in the Ta-strip. The **h**$_{EFF}$ takes into account the standard micromagnetic energy contributions from external, magnetostatic and exchange field, the Oersted field from both $J_{Ta}$ and $J_{MTJ}$, and the dipolar coupling from the pinned layer. First of all, we carried out preliminary numerical simulations of the same structure studied by Liu *et al.* in Ref. 10, analysing different cross-sections and free layer thicknesses, in order to geometrically optimize the device response in terms of magnetization dynamics. In the following, we present the micromagnetic study for the geometry configuration where we obtained large amplitude magnetization precession. In this case, the dimensions of the ellipse are: $w$=100 nm along the *x*-axis, $l$=300 nm along the *y*-axis, and thickness $t$=2 nm. The advantage to use a larger thickness (in the structure by Liu *et al.*[10] it was 1.5 nm) consists in a better understanding of the STT effect, being the perpendicular anisotropy and the Dzyaloshinskii–Moriya interaction negligible.[3, 12, 22] Particularly, we identified a configuration which permits to excite a quasi-uniform mode and to achieve promising results for the injection locking phenomenon driven by a "weak" microwave STT and a fixed bias $J_{Ta}$. The used physical parameters are: saturation magnetization $M_S$=1000x10$^3$ A/m, exchange constant $A$=2.0x10$^{-11}$ J/m, magnetic damping α=0.015, spin-hall angle $α_H$ =-0.15, and spin-polarization $\eta_T$ =0.66.

Fig. 2a shows the oscillation frequency as a function of $J_{Ta}$ related to the oscillation of the *y*-component of the free layer magnetization for two different field amplitudes $H_{ext}$=30 and 40 mT ($J_{MTJ}$=0 A/cm$^2$). The external field is applied with an in-plane angle tilted $\phi$=30° with respect to the *x*-axis of the ellipse. For this thickness, the critical current densities are of the order of 10$^8$ A/cm$^2$ and are almost independent on the field amplitude (at least for the simulated values 20-50 mT).

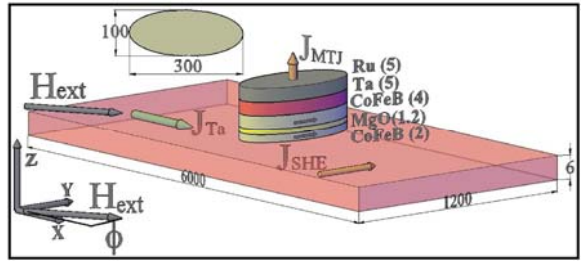

FIG. 1 (color online) Schematic representation of the three terminal MTJ device.

As the field amplitude increases, a decreasing of the current region where coherent magnetization dynamics is observed (e.g. at $H_{ext}$=40 and 30 mT the current range are between -1.65 and -1.95x10$^8$A/cm$^2$ and between -1.38 and -1.97x10$^8$ A/cm$^2$ respectively).

As expected, the oscillation frequency increases with the amplitude of the external field and its value at the critical current is 3.75 GHz for $H_{ext}$ =30 mT and 4.60 GHz for $H_{ext}$ =40 mT. The oscillation frequency exhibits a slightly red-shift as function of $J_{Ta}$, indicating the presence of an in-plane oscillation axis, as also confirmed by the temporal evolution of the magnetization. Fig. 2b shows the normalized average components of the magnetization $<m_x>$, $<m_y>$, $<m_z>$ (dashed, solid, and dotted line respectively) for $J_{Ta}$=-2.13 x 10$^8$ A/cm$^2$ and $H_{ext}$=30 mT for a time of 1 ns. In this case, a large amplitude of the oscillation mode in the *x-y* plane is shown. The spatial configuration of the magnetization snapshots are displayed in Fig. 2c (red positive and blue negative *y*-component of the magnetization [see supplementary material movie 1]), related to the numbers 1-6 as displayed in Fig.2b. Quasi-uniform magnetization dynamic is observed, with the magnetization oscillating of 180 degrees back and forth (compare snapshot 1 and 4). In other words, the *y*-component of the magnetization rotates firstly clockwise (points 1-4) and then counter-clockwise (points 4-6).

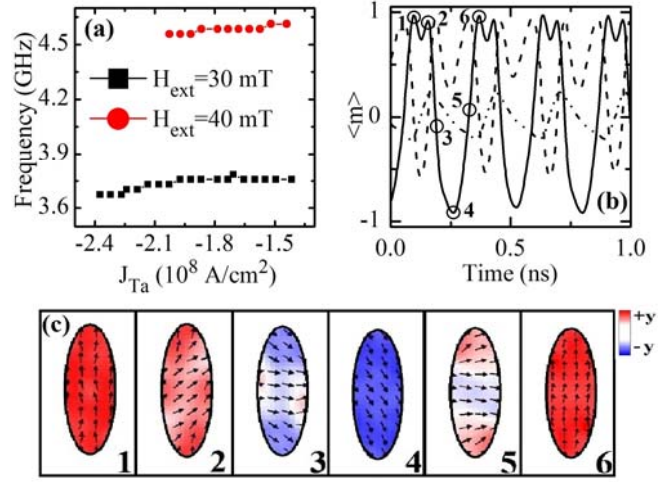

FIG. 2 (color online) (a) Oscillation frequency of the magnetization as a function of the $J_{Ta}$ for $H_{ext}$=40 mT (top curve) and $H_{ext}$=30 mT (bottom curve) when the $J_{MTJ}$ is zero. (b) Temporal evolution of the three normalized components of the magnetization $<m_x>$ (dashed curve), $<m_y>$ (solid curve), $<m_z>$ (dotted curve) during 1 ns of the magnetization oscillations, for $J_{Ta}$=-2.13x10$^8$ A/cm$^2$, $H_{ext}$=30 mT. (c) Snapshots of the magnetization during an oscillation period in the time instants reported in Fig. 2b. The color scale refers to the y-component of the magnetization (red positive, blue negative). The arrows indicate the magnetization direction. (enhanced online).

Fig. 3a shows the oscillation frequency of the main excited mode for a fixed $J_{Ta}$=-2.13x10$^8$ A/cm$^2$ as function of a bias $J_{MTJ}$. For positive $J_{MTJ}$, the oscillation frequency exhibits small variation near 3.75 GHz, while a large frequency tunability around 100 MHz/(10$^6$ A/cm$^2$) for negative $J_{MTJ}$ is observed. This result can be explained in the following way. A positive $J_{MTJ}$ acts as an additional positive damping, in fact, for $J_{MTJ}$ larger than 7x10$^6$ A/cm$^2$, the microwave emission is switched off. On the contrary, a negative $J_{MTJ}$ acts as a negative damping, showing a significant role in the oscillator frequency.[23] This behavior is different from the one observed experimentally,[9] where a linear tunability of the oscillation frequency on current was found with both signs of the $J_{MTJ}$. Indeed, in that particular framework, the $J_{MTJ}$ was used to modify the perpendicular anisotropy, whereas in this case the anisotropy contribution is negligible. Fig. 3b shows the Fourier spectra for different values of the $J_{MTJ}$. In agreement with the experimental data, it can be observed that the amplitude of the peak increases with increasing the negative value of the current.

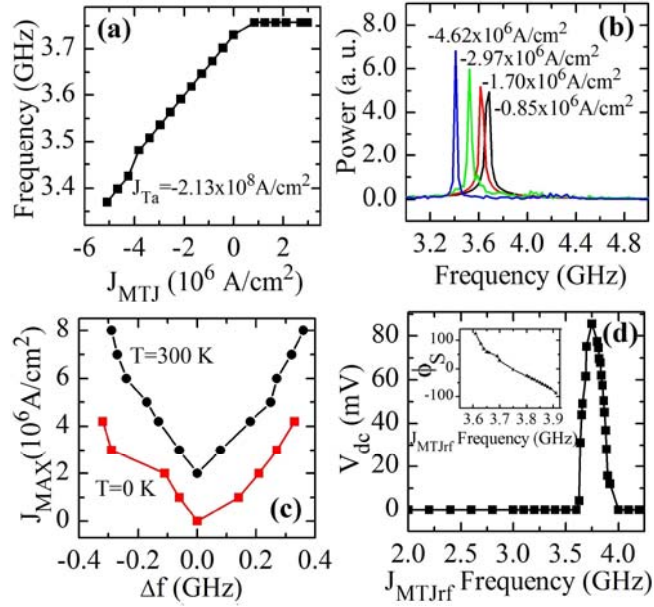

FIG. 3 (color online) (a) Oscillation frequency for fixed $H_{ext}$=30 mT and $J_{Ta}$ =-2.13 x $10^8$ A/cm$^2$ as function of the bias $J_{MTJ}$. (b) Fourier spectra for different values of the $J_{MTJ}$. (c) Arnold tongues showing the locking regions as function of $J_{MTJrf}$ for $T$=0 K (lower curve) and $T$=300 K (upper curve) at $J_{Ta}$=-2.13 x $10^8$ A/cm$^2$. (d) Dc output voltage vs $f_{rf}$ as computed with Eq. 3 for $J_{MAX}$ =2x$10^6$ A/cm$^2$. Inset: intrinsic phase shift $\phi_S$ as function of $f_{rf}$ inside the locking region ($J_{MAX}$ =2x$10^6$ A/cm$^2$).

One of the main properties of STOs is the possibility to control the output frequency of the self-oscillation via the injection locking phenomenon.[24, 25, 26, 27] For in-plane magnetized free layer, the injection locking has been observed at the 2$^{nd}$-harmonic (in our case the frequency of the y-component of the magnetization).[28] In general, the microwave currents were applied to the same terminal of the bias current.[29] Here, the magnetization precession is driven by the $J_{Ta}$, while the injection locking is due to a microwave current density $J_{MTJrf} = J_{MAX}\text{sen}(2\pi f_{rf} t)$ applied via the third terminal ($J_{MAX}$ and $f_{rf}$ are the amplitude and the microwave frequency). In other words, this system permits to study the non-autonomous behavior of an STO by separating electrically the biasing current from the microwave source. We fixed $J_{Ta}$=-2.13 x $10^8$ A/cm$^2$ and $H_{ext}$=30 mT, which corresponds to an oscillation frequency of 3.75 GHz. The locking properties have been studied for a $J_{MTJrf}$ with amplitude $J_{MAX}$ from 1 to 4.2x$10^6$ A/cm$^2$ at $T$=0 K and up to 8x$10^6$ A/cm$^2$ for $T$=300 K and a microwave frequency from 3.0 GHz to 8.0 GHz. Fig. 3c summarizes the locking range $\Delta$ as function of $J_{MAX}$, without and with the thermal fluctuations ($T$=300 K).[30, 31] For example, at $J_{MAX}$=2x$10^6$ A/cm$^2$, the locking range is $\Delta$=320 MHz, from 3.61 to 3.93 GHz. For current densities up to 4.2x$10^6$ A/cm$^2$ the response is qualitatively the same, the $\Delta$ increases linearly with $J_{MAX}$. The presence of thermal fluctuations imposes a larger $J_{MAX}$ to reach the same locking region $\Delta$. No qualitative differences are observed by changing -0.1<$\alpha_H$<-0.2 and 0.5<$\eta_T$<0.7, and by considering different MTJ cross sections of 310x100 nm$^2$ and 290x100 nm$^2$. Our results predict locking regions comparable or even larger than the experimental ones for microwave current densities of the same order.[23, 24, 28] In the synchronization region, where the resistance $r$ oscillates at the same frequency $\omega_S$ of the microwave source, the signal can be written as $r=R_{<M>,S}+\Delta R_S \sin(\omega_S t+\phi_S)$, being $\Delta R_S$ and $R_{<M>,S}$ the amplitude of the oscillating tunnelling magnetoresistive signal and its mean value respectively, and $\phi_S$ the intrinsic phase shift in the synchronized state.[32] The output voltage $v_0$ over the MTJ is given by:

$$v_0 = \left(R_{<M>,S}+\Delta R_S \sin(\omega_S t+\phi_S)\right)I_{MAX}\sin(\omega_S t) =$$
$$= R_{<M>,S}I_{MAX}\text{sen}(\omega_S t)+\frac{\Delta R_S I_{MAX}}{2}(\cos(\phi_S)-\cos(2\omega_S t+\phi_S)) \quad (2)$$

where $I_{MAX}=SJ_{MAX}$ ($S$ is the cross section of the free layer). Together to the microwave signals at $2\omega_S$ and $\omega_S$ that can be used for the design of microwave oscillators, a dc component $0.5\Delta R_S I_{MAX}\cos\phi_S$ is also observed. Fig. 3d shows the dc output voltage as function of the microwave frequency for a $J_{MAX}$=2x$10^6$ A/cm$^2$ ($R_P$=4450 $\Omega$ and $R_{AP}$=5200 $\Omega$). A maximum voltage of 80 mT is achieve inside $\Delta$, whereas zero dc voltage is measured outside $\Delta$. The inset of Fig. 3d shows the intrinsic phase shift $\phi_S$ as function of the microwave frequency. The prediction of this large dc voltage makes this system very promising for the design of a next generation of high sensitive resonant microwave signal detectors.[33] The results described above are at the basis of the scalable synchronization scheme discussed below.

Now, we focus our attention to MTJs with different cross sections (MTJ$_1$, MTJ$_2$, and MTJ$_3$) with in plane axes 310 and 100 nm, 300 and 100 nm (same studied above), 290 and 100 nm respectively. Fig. 4a shows a sketch of the proposed synchronization scheme for the three MTJs, but we stress the fact that this system is highly scalable and it can be extended to an array of $N$-three terminal systems.

The dependence of the oscillation frequency on $J_{Ta}$ in MTJ$_1$ and MTJ$_3$ is similar to the one related to the MTJ$_2$ (not shown). For a fixed $J_{Ta}$, the locking range of the three MTJs is of the same order, but centered over a different oscillation frequency. For example, at $J_{Ta}$=-2.13x10$^8$ A/cm$^2$ and $J_{MAX}$=2.0x10$^6$ A/cm$^2$, we achieved for MTJ$_1$ a magnetization precessional frequency $f_1$=3.62 GHz and $\Delta_1$=390 MHz, for MTJ$_2$ $f_2$=3.75 GHz and $\Delta_2$=320 MHz, and for MTJ$_3$ $f_3$=3.78 GHz and $\Delta_3$=310 MHz. As illustrated in Fig. 4b, the locking ranges are overlapped for a region of 290 MHz, suggesting a possible way to synchronize parallel connected three terminal oscillators.

The magnetization precession is excited by means of the SHE in all the MTJs. The synchronization is achieved via a microwave voltage applied to the third terminal $V_{RF}=V_M \sin(\omega_S t)$. The output signal can be read as the voltage over $R_0$. For each $i$-MTJ, the conductance $G_i$ is given by $G_i = G_{<M>,i} + \Delta G_i \sin(\omega_i t + \phi_i)$ where $G_{<M>,i}$ is the average conductance when the magnetization precession is excited, $\Delta G_i$, $\omega_i$ and $\phi_i$ are, respectively, the amplitude, the frequency, and the intrinsic phase shift[32] of the oscillation generated in the $i$-MTJ. For $N$-synchronized MTJs at the frequency $\omega_S$. The total conductivity is given by $G_T = \sum_{i=1...N} G_{<M>,i} + \sum_{i=1...N} \Delta G_i \sin(\omega_S t + \phi_i)$. The electrical circuit is completed by adding two filters with the aim to use the synchronization scheme to enhance the output microwave power at $2\omega_S$ or the dc voltage. In the case of pass-band filters, $Z_0$ and $Z_1$ are composed by a capacitor and an inductor connected in series, where $L_0 C_0 = \frac{1}{4\omega_S^2}$ and $L_1 C_1 = \frac{1}{\omega_S^2}$ respectively. In this way, the output voltage over $R_0$ is given by:

$$v_0 = R_0 i_0 = \frac{R_0 V_M}{2} \sum_{i=1...N} \Delta G_i \cos(2\omega_S t + \phi_i) \qquad (3)$$

When $\phi_i$ values are the same (or within a range smaller than 10 degree), $v_0$ can be approximated to the sum of the signals from the MTJs as $v_0 \approx 0.5 R_0 V_M \cos(2\omega_S t + \phi_S) \sum_{i=1...N} \Delta G_i$. If $Z_0$ is a low pass filter (i.e. a capacitor) and $Z_1$ is a high pass filter (i.e. an inductor), the output voltage over $R_0$ is given by:

$$v_0 = -\frac{R_0 V_M}{2} \sum_{i=1...N} \Delta G_i \cos(\phi_i) \qquad (4)$$

The equations (3) and (4) point out how the proposed synchronization scheme can give rise to an improvement of the dynamical properties (for example power) if used as oscillator, or to an enhancement of the sensitivity (output dc voltage over the power of the microwave signal) when used as microwave detector.

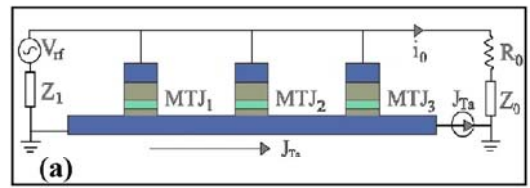

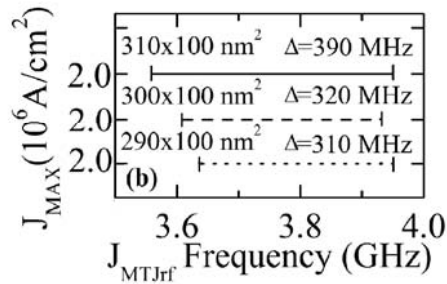

FIG. 4 (a) Schematic representation of the proposed highly scalable synchronization scheme. (b) Locking ranges as a function of the cross-section dimensions for $J_{MAX}$=2.0 x $10^6$ A/cm$^2$ and $J_{Ta}$ =-2.13 x $10^8$ A/cm$^2$.

In summary, we have micromagnetically studied the dynamical behavior of a three terminal MTJs driven by the SOT and the STT. We have found that the control of the STT and the SOT via electrically separated terminals opens promising perspective from a technological point of view in the design of next generation of spintronic oscillators and microwave detectors, overcoming the limit of the output power and sensitivity by means of an innovative highly scalable synchronization scheme.

This work was supported by project MAT2011-28532-C03-01 from Spanish government, and project PRIN2010ECA8P3 from Italian MIUR.